\documentclass[aps,prl,showpacs,twocolumn,superscriptaddress,nofootinbib,floatfix,amsmath,10pt]{revtex4}
\usepackage{epsfig,epsf,graphics,psfrag}
\usepackage{times}
\usepackage{color}
\usepackage{amsfonts,amssymb,stmaryrd,latexsym,amsmath}
\usepackage{hhline}

\newcommand{\be}{\begin{equation}}
\newcommand{\ee}{\end{equation}}
\newcommand{\ba}{\begin{eqnarray}}
\newcommand{\ea}{\end{eqnarray}}

\begin{document}
\title{Viability of Carbon-Based Life as a Function of the Light Quark Mass}

\author{Evgeny Epelbaum}
\affiliation{Institut~f\"{u}r~Theoretische~Physik~II,~Ruhr-Universit\"{a}t~Bochum,
D-44870~Bochum,~Germany}

\author{Hermann~Krebs}
\affiliation{Institut~f\"{u}r~Theoretische~Physik~II,~Ruhr-Universit\"{a}t~Bochum,
D-44870~Bochum,~Germany}

\author{Timo~A.~L\"{a}hde}
\affiliation{Institut~f\"{u}r~Kernphysik, Institute~for~Advanced~Simulation, and
J\"{u}lich~Center~for~Hadron~Physics,~Forschungszentrum~J\"{u}lich,
D-52425~J\"{u}lich, Germany}

\author{Dean~Lee}
\affiliation{Department~of~Physics, North~Carolina~State~University, Raleigh, 
NC~27695, USA}

\author{Ulf-G.~Mei{\ss }ner}
\affiliation{Institut~f\"{u}r~Kernphysik, Institute~for~Advanced~Simulation, and
J\"{u}lich~Center~for~Hadron~Physics,~Forschungszentrum~J\"{u}lich,
D-52425~J\"{u}lich, Germany}
\affiliation{Helmholtz-Institut f\"ur Strahlen- und
             Kernphysik and Bethe Center for Theoretical Physics, \\
             Universit\"at Bonn,  D--53115 Bonn, Germany}
\affiliation{JARA~-~High~Performance~Computing, Forschungszentrum~J\"{u}lich, 
D-52425 J\"{u}lich,~Germany}

\begin{abstract}
\noindent
The Hoyle state plays a crucial role in the helium burning 
of stars that have reached the red giant stage.  
The close proximity of this state to the
triple-alpha threshold is needed for the production of carbon, oxygen, 
and other elements necessary for life.  We investigate whether this
life-essential condition is robust or delicately fine-tuned by
measuring its dependence on the fundamental constants of nature,
specifically the light quark mass and the
strength of the electromagnetic interaction. We show that there
exist strong correlations between the alpha-particle binding energy 
and the various energies relevant
to the triple-alpha process.
We derive limits on the variation of these fundamental 
parameters from the requirement that sufficient
amounts of carbon and oxygen be generated in stars. We also discuss the
implications of our results for an anthropic view of the Universe.
\end{abstract}

\maketitle  

Life as we know it depends on the availability of carbon and oxygen.
These two essential elements are produced during helium burning in red 
giant stars. The initial reaction is the so-called triple-alpha process, 
where three helium nuclei fuse to generate $^{12}$C. This can be viewed 
as a two-step process. First, two $^4$He nuclei combine to form an unstable, 
but long-lived $^8$Be resonance. This $^8$Be resonance must then combine 
with a third alpha-particle to generate carbon. By itself, this process 
cannot explain the observed abundance of carbon in 
the Universe. Therefore, Hoyle postulated that a new excited state of
$^{12}$C, a spinless even-parity resonance near 
the $^{8}$Be-alpha threshold, enhances the reaction~\cite{1}. Soon after 
this prediction, the new state was found at 
Caltech~\cite{2,3} and has since been investigated in laboratories worldwide. 
The measured energy of this second $0^+$ state is $\varepsilon =
379.47(18)$~keV above the triple-alpha threshold, while the total and radiative 
widths are known to be $\Gamma_{\rm tot}^{} = 8.3(1.0)$~eV and
$\Gamma_{\gamma}^{} = 3.7(5)$~meV, respectively. The reaction rate for the 
(resonant) triple-alpha process is approximately given by~\cite{Oberhummer:2000mn}
\begin{equation}
\label{eq:rate}
r_{3\alpha}^{} \propto \Gamma_\gamma^{} \, (N_\alpha^{}/k_{B}^{}T)^3 
\exp(-\varepsilon/k_{B}^{}T),
\end{equation}
with $N_\alpha^{}$ the alpha-particle number density, $T$ the stellar
temperature and $k_B^{}$ Boltzmann's constant. Due to the exponential
dependence, $\varepsilon$ is the dominant control parameter of this reaction. 
Here, we study the dependence of $\varepsilon$ upon the fundamental 
parameters of the strong and electromagnetic (EM) interactions.

Given its role in the formation of life-essential elements, the Hoyle state
has been called the ``level of life''~\cite{Linde} (see Ref.~\cite{Kragh} for a
thorough discussion of the history of this issue). 
Thus, it is often considered a prime example of the anthropic principle, which states 
that the observed values of the fundamental physical and cosmological
parameters are restricted by the requirement 
that life can form to observe them, and that the current Universe be old 
enough for that to happen~\cite{6,7}. 
In the context of cosmology and string theory, consequences derived from 
anthropic considerations have had considerable impact 
(see {\it e.g.} Refs.~\cite{Weinberg:1987dv,Susskind:2003kw}).

Several numerical studies have investigated the impact of changes in the
Hoyle state 
energy. Livio {\it et al.}~\cite{Livio} modified the 
value of $\varepsilon$ by hand and performed calculations involving the 
triple-alpha process in the core and helium shell burning of helium up to
the asymptotic giant branch stage in stellar evolution. They concluded that 
a $\simeq 60$~keV change in $\varepsilon$ could be tolerated, 
and thus the amount of fine-tuning required was not as severe as first believed.  

A more microscopic calculation was performed by Oberhummer 
{\it et al.}~\cite{Oberhummer:2000mn,Oberhummer:2000zj}
in terms of a nuclear cluster model based on a simple two-nucleon (NN) + EM
interaction. This NN interaction was formulated in terms of one strength 
parameter, adjusted to give a fair description of $\alpha$--$\alpha$ scattering
and the spectrum of $^{12}$C. By modifying this coupling strength and the
EM fine structure constant $\alpha_{\rm em}^{}$, the effect on 
carbon and oxygen production was analyzed.
Outside of a narrow window of $\simeq 0.5$\% 
around the observed strong force and $\simeq 4$\% around the observed Coulomb 
force, the stellar production of carbon and/or oxygen was found to be reduced 
by several orders of magnitude. However, this model 
of the strong force is not readily connected to the fundamental theory of 
the strong interactions, quantum chromodynamics (QCD), and its 
fundamental parameters, the light quark masses. Therefore, it is not obvious 
how to translate the findings of Ref.~\cite{Oberhummer:2000zj} into anthropic 
constraints on fundamental parameters. In this study, we shall address 
this pertinent question: What changes in the quark masses and the
EM fine structure constant are 
consistent with the formation of carbon-based life? 

Over the last few years, we have developed a new method to study 
atomic nuclei and their properties from first principles, termed 
{\sl nuclear lattice simulations}. The key ingredients in this approach are,
on the one hand, the chiral effective field theory (EFT)
of nuclear forces and, on the other hand, large-scale lattice Monte 
Carlo methods. The latter are also fruitfully used in many other fields 
of science. Chiral nuclear EFT was
introduced by Weinberg~\cite{Weinberg:1990rz} (for a first numerical implementation, see \cite{Ordonez:1995rz}) as a systematic tool
to explore the consequences of spontaneous and explicit
chiral symmetry breaking of QCD in a rigorous manner.  The basic degrees of
freedom are pions and nucleons, where the pions and their interactions
carry the basic information of the chiral symmetry properties of QCD.
In particular, one finds $M_\pi^2 \sim (m_u^{}+m_d^{})$, so that any dependence
on the light quark masses $m_u^{}$ and $m_d^{}$ can be translated into a 
corresponding dependence on the pion mass $M_\pi^{}$. In what follows, 
only the average light quark mass $m_q \equiv (m_u^{}+m_d^{})/2$ will be considered, 
as the effects of strong isospin violation due to $m_u^{} \neq
m_d^{}$ are greatly suppressed for the reactions considered here.
Chiral nuclear EFT is based on an order-by-order expansion of 
the nuclear potential. In this scheme, two-, three- and four-nucleon forces arise 
naturally, and 
their observed hierarchy 
is also  explained. 
The nuclear forces have been worked out to high precision and applied successfully
in few-nucleon systems for binding energies, structure, and reactions.  
For recent reviews, see Refs.~\cite{Epelbaum:2008ga,Machleidt:2011zz}. 
Within chiral nuclear EFT, the quark mass dependence of light nuclei and 
its impact on big bang nucleosynthesis has already been studied; see, {\it e.g.},
Refs.~\cite{Beane:2002vs,Epelbaum:2002gb,Braaten:2003eu,Bedaque:2010hr,Chen:2010yt,Soto:2011tb}
and Ref.~\cite{Flambaum:2007mj} for a related study.

Monte Carlo simulations have been used to solve the nuclear
$A$--body problem (with $A$ the atomic number) based on a 
lattice formulation~\cite{Lee:2004si}. The lattice spacing of this discretized space-time
serves as an ultraviolet 
regulator. The nucleons are placed on the lattice sites, and the
interactions are represented by pionic and (suitably chosen)
auxiliary fields.
Our periodic cubic lattice has
a spacing of $a = 1.97$ fm and a length of $L = 11.82$ fm. In
the time direction, our lattice spacing is $a_t^{} = 1.32$ fm, and the
propagation time $L_t^{}$ is varied in order to extrapolate to $L_t^{} \to \infty$. 
The energies of the ground and excited states are obtained using projection
Monte Carlo techniques~\cite{Epelbaum:2011md,Epelbaum:2012qn}. 
More precisely, we compute
$Z_A^{}(t) \equiv \langle \psi_A^{}| \exp(-Ht) |\psi_A^{}\rangle$
for a given $A$--nucleon system at large Euclidean time $t$
in order to extract the energies of the low-lying states; see also
Ref.~\cite{Lee:2008fa} for more details.   

The leading order (LO) contribution to the NN force 
emerges from the one-pion exchange potential (OPEP) and (smeared) $S$--wave contact
interactions. This improved LO action forms the basis of our projection Monte Carlo
simulations, while all higher-order terms including the 
Coulomb interaction, corrections to the NN force and 
three-nucleon forces, are treated in perturbation theory. 
All parameters of $H$ are fixed from two- and three-nucleon
data, enabling predictions for all heavier nuclei. So far, such
calculations have been performed up to next-to-next-to-leading 
order (NNLO), achieving a good description of nuclei
up to $A = 12$. 


We have performed the first {\it ab initio} calculations for the  
energy~\cite{Epelbaum:2011md} and structure of the Hoyle state~\cite{Epelbaum:2012qn} 
using the nuclear lattice formalism (see 
Ref.~\cite{Navratil:2007we} for a no-core shell model calculation of the spectrum of $^{12}$C
employing chiral EFT forces). In our approach, 
the hadronic interactions of the nucleons with themselves and with pions 
can be modified 
easily. Our analysis of the dependence 
upon the strength of the Coulomb interaction is 
therefore straightforward.  
For the 
dependence on $m_q^{}$, we also need information about the 
quark mass dependence of the hadronic interactions.  
In turn, such dependences can be given as a function of $M_\pi^{}$.

We shall restrict ourselves to values of $M_\pi^{}$ near the physical point, 
with $|\delta M_\pi^{}/M_\pi| \leq 10\%$. Such small changes can be treated in
perturbation theory.
The $M_\pi$--dependence of the OPEP
and
the nucleon mass $m_N^{}$ is determined in chiral perturbation theory utilizing
constraints from lattice QCD, see Ref.~\cite{Kfactors} for more
details. To retain model independence, we do not rely on the chiral
expansion of the NN contact interactions. Instead, we 
express our results in terms of the derivatives of the inverse
spin-singlet and spin-triplet NN scattering lengths
with respect to the pion mass,
\begin{equation}\label{eq:deriv}
\bar{A}_s^{} \equiv \left. \frac{\partial a_s^{-1}}{\partial M_\pi^{}}
\right|_{M_\pi^\mathrm{ph}}, \quad
\bar{A}_t^{} \equiv \left. \frac{\partial a_t^{-1}}{\partial M_\pi^{}} \right|_{M_\pi^\mathrm{ph}},
\end{equation}
which parameterize the $M_\pi^{}$--dependence of the
short-range nuclear force and can be measured in lattice QCD. 
We do not consider $M_\pi^{}$--dependent short-range
effects beyond the ones introduced above. (The correlations observed
for various energy differences, as discussed below, indicate that the 
dynamics of interest is largely governed by the large $S$--wave NN
scattering lengths. Higher-order $M_\pi^{}$--dependent short-range terms
are therefore expected to play a minor role.) 
Thus, 
the variation of a given nuclear 
energy level $E_i^{}$ takes the form
\begin{align} 
\label{eq:E}
\left. \frac{\partial E_i^{}}{\partial M_\pi^{}} \right|_{M_\pi^{\rm ph}} & =
\left. \frac{\partial E_i^{}}{\partial M_\pi^\mathrm{OPE}}
\right|_{M_\pi^\mathrm{ph}} 
\!\! + x_1^{} \left. \frac{\partial E_i^{}}{\partial m_N^{}} \right|_{m_N^\mathrm{ph}}
\\
& + x_2^{} \left. \frac{\partial E_i^{}}{\partial \tilde g_{\pi N}^{}}
\right|_{\tilde g_{\pi N}^\mathrm{ph}}
\!\! + x_3^{} \left. \frac{\partial E_i^{}}{\partial C_{0}^{}} \right|_{C_{0}^\mathrm{ph}}
\!\! + x_4^{} \left. \frac{\partial E_i^{}}{\partial C_{I}^{}} \right|_{C_{I}^\mathrm{ph}},
\nonumber
\end{align}
with $x_1^{} \equiv \partial m_N^{}/\partial M_\pi^{}|_{M_\pi^{\rm ph}}$, 
$x_2^{} \equiv \partial \tilde{g}_{\pi N}^{}/\partial M_\pi^{}|_{M_\pi^{\rm ph}}$ {\it etc.}, where
$X^{\rm ph}_{}$ denotes the value of $X$ for the physical $M_\pi^{}$. 
The terms in Eq.~(\ref{eq:E}) represent different contributions to the pion mass
variation. First, there is the explicit dependence on $M_\pi^{}$ through the pion
propagator in the 
OPEP. 
Second, we include the dependences on $M_\pi^{}$ through the nucleon mass $m_N^{}$ and 
${\tilde g}_{\pi N}^{} \equiv g_A^{}/(2F_\pi^{})$, with $g_A^{}$ the nucleon axial-vector coupling and $F_\pi^{}$ the weak pion decay constant.  
Finally, we have the $M_\pi^{}$--dependences 
from the strengths of the NN contact interactions $C_0^{}$ and 
$C_I^{}$, which are expressed through the derivatives given in 
Eq.(\ref{eq:deriv}). Therefore, the problem reduces 
to the calculation of various derivatives of the nuclear energy levels using lattice 
Monte Carlo techniques and the determination of the coefficients $x_1^{}\ldots x_4^{}$.
The derivatives of $E_i^{}$ in Eq.~(\ref{eq:E}) are computed by evaluating the 
expectation value of the derivative of the lattice Hamiltonian $H$ with respect to $M^\mathrm{OPE}_{\pi}$, 
$m_N^{}$, $\tilde g_{\pi N}^{}$, $C_0^{}$ and $C_I^{}$. This involved the generation of $\mathcal{O}(10^7)$ 
statistically independent pion- and auxiliary field configurations on the Blue Gene/Q supercomputer JUQUEEN using
the hybrid Monte Carlo algorithm. The explicit form of $H$ can be found in Ref.~\cite{EPJA_long}.

The values of $x_1^{}$ and $x_2^{}$ can be obtained from lattice QCD combined
with chiral extrapolations (see, {\it e.g.}, Ref.~\cite{Kronfeld:2012uk} for a recent 
review on lattice QCD and determinations of the nucleon mass variation). 
We exchange $x_3^{}$ and $x_4^{}$ for $\bar{A}_s^{}$ and $\bar{A}_t^{}$ by consideration
of the $M_\pi^{}$--dependence of NN scattering in a cubic box.
We may then compute the energy differences $\Delta E_h^{} \equiv E_{12}^* - E_8^{} -
E_4^{}$ and $\Delta E_b^{} \equiv E_8^{} - 2E_4^{}$, where $E_{12}^*$ is the energy of
the Hoyle state and $E_{4,8}^{}$ the ground-state energies of the
$^4$He and $^8$Be nuclei, respectively.
Note also that $\varepsilon \equiv \Delta E_h^{} + \Delta E_b^{}$.
We find
\begin{align}
\left. \frac{\partial \Delta E_h^{}}{\partial M_\pi^{}}
\right|_{M_\pi^\mathrm{ph}} 
\hspace{-.2cm} & = 
- 0.455(35) \bar{A}_s^{}
-0.744(24) \bar{A}_t^{}
+0.051(19), \nonumber \\
\left. \frac{\partial \Delta E_b^{}}{\partial M_\pi^{}}
\right|_{M_\pi^\mathrm{ph}}  
\hspace{-.2cm} & =
-0.117(34) \bar{A}_s^{} -0.189(24) \bar{A}_t^{}
+0.013(12), \nonumber \\
\left. \frac{\partial \varepsilon}{\partial M_\pi^{}} \right|_{M_\pi^\mathrm{ph}} 
\hspace{-.2cm} & =
-0.572(19) \bar{A}_s^{} -0.933(15) \bar{A}_t^{}
+0.064(16),
\label{MC_res}
\end{align}
where the parentheses represent the one-standard-deviation
stochastic and extrapolation error, combined with the uncertainty in 
$x_{1,2}^{}$ 
(as explained in Ref.~\cite{Kfactors}) which affects only the constant (OPEP) terms above. 

\begin{figure}[t!]
\begin{center}
\vspace{-2.5mm}
\includegraphics[width=.87\columnwidth]{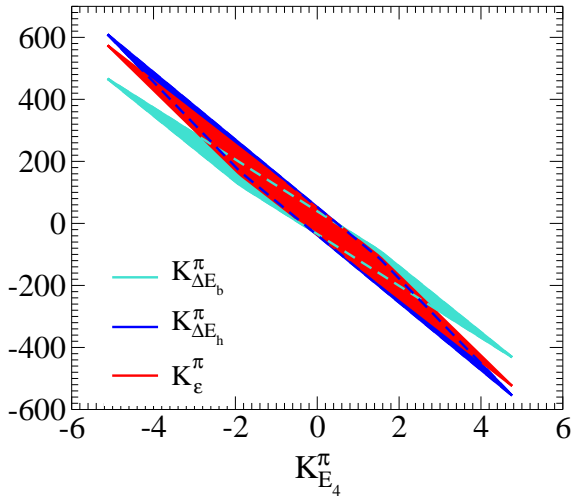}
\end{center}
\par
\vspace{-0.4cm}\caption{Correlation of the $\alpha$-particle energy $E_4^{}$ with the
energy differences pertinent to the triple-alpha process 
(the bands correspond to $\Delta E_b^{}$, $\varepsilon$, and $\Delta E_h^{}$ in clockwise order)
under variation of
$\bar{A}_{s,t}^{}$ in the range $\{-1 \ldots 1\}$.
}
\label{fig:correlations}
\end{figure}

The results in Eq.~(\ref{MC_res})
are intriguing. First, we note that $(\partial \Delta E_h^{}/\partial M_\pi^{}) 
/ (\partial \Delta E_b^{}/\partial M_\pi^{}) \simeq 4$, thus $\Delta E_h^{}$
and $\Delta E_b^{}$ cannot be independently fine-tuned.  Such behavior can 
readily be explained in terms of the $\alpha$--cluster structure of the Hoyle state and $^8$Be. 
Further correlations are visualized in Fig.~\ref{fig:correlations}, where
the relative changes in $\Delta E_b^{}, \Delta E_h^{}$ and $\varepsilon$ are shown as a 
function of relative changes in the ground state energy $E_4^{}$ of the $\alpha$-particle.
We define $K^\pi_X \equiv (\partial X / \partial M_\pi^{}) M_\pi^{}/X$ as the relative 
variation of $X$ with respect to $M_\pi^{}$.
Fig.~\ref{fig:correlations} provides clear evidence that the alpha binding energy 
is strongly correlated with $\Delta E_b^{}$, $\Delta E_h^{}$, and $\varepsilon$. 
Such correlations related to carbon production have been speculated upon 
earlier~\cite{Livio,WeinbergFacing}.  
Second, we note that there is a special value for the ratio of $\bar{A}_s^{}$ to $\bar{A}_t^{}$,
given by
\begin{equation}
\label{eq:magic}
\bar{A}_s^{} / \bar{A}_t^{} 
\simeq -1.5~,
\end{equation}
where the pion mass dependence of $\Delta E_h^{}$, $\Delta E_b^{}$, and $\varepsilon$ 
becomes small (compared to the error bars). 

We have expressed all our results in terms of the quantities
$\bar{A}_{s,t}^{}$, the quark mass dependence of which was considered 
at next-to-leading order (NLO) in Ref.~\cite{Epelbaum:2002gb}, with a recent
update to NNLO~\cite{Kfactors}.
That analysis gives $\bar{A}_s^{} = 0.29^{+0.25}_{-0.23}$ and $\bar{A}_t^{} 
= -0.18^{+0.10}_{-0.10}$, where the errors reflect the theoretical 
uncertainties. As expected, these values of $\bar{A}_{s,t}^{}$ are of 
natural size. Taking into account correlations
in the calculation of $\bar{A}_{s,t}^{}$, we find $\bar{A}_s^{}/\bar{A}_t^{} 
= -1.6^{+1.0}_{-1.7}$. Interestingly, the central value is very close to the 
result given in Eq.~(\ref{eq:magic}), for which the 
pion mass dependences of $\Delta E_h^{}$, $\Delta E_b^{}$, and $\varepsilon$ 
are all approximately zero (within error bars). In the future, a reduction of the uncertainty
in $\bar{A}_{s,t}^{}$ is desirable. This can be addressed by lattice QCD calculations of 
NN systems. For recent studies, see Refs.~\cite{Beane:2006mx,Yamazaki:2012hi}.

\begin{figure}[t!]
\begin{center}
\vspace{-2.5mm}
\includegraphics[width=.87\columnwidth]{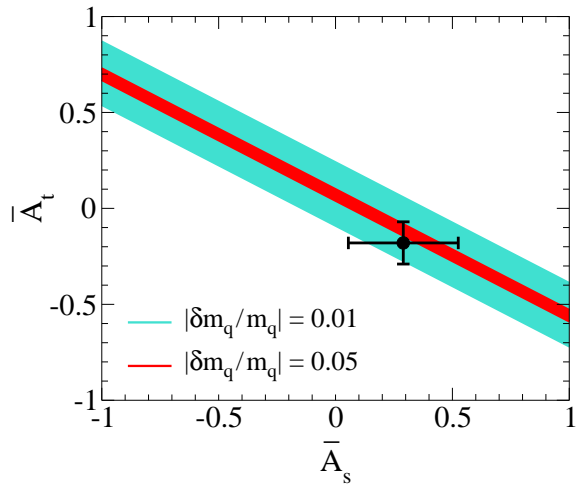}
\end{center}

\vspace{-0.3cm}

\caption{``Survivability bands'' for carbon-oxygen based life from Eq.~(\ref{MC_res}), due to 
1\% (broad outer band) and 5\% (narrow inner band) changes in $m_q^{}$
in terms of the parameters 
$\bar{A}_s^{}$ and $\bar{A}_t^{}$. The most up-to-date NNLO analysis of $\bar{A}_{s,t}^{}$
is depicted by the data point with horizontal and vertical error bars.}
\label{fig:endoftheworld}
\end{figure}

We now use the reaction rate in Eq.~(\ref{eq:rate})
to draw conclusions about the allowed variations of the fundamental constants. 
From the stellar modeling calculations in Ref.~\cite{Oberhummer:2000zj}, we find that
sufficient abundances of both carbon and oxygen can be maintained within an envelope 
of $\pm 100$~keV around the observed value of $\varepsilon$. Allowing for a maximum shift of 
$\pm 100$~keV in $\varepsilon$ translates into bounds on the
variations of 
$m_q^{}$. 
In Fig.~\ref{fig:endoftheworld}, we show 
``survivability bands'' for carbon-oxygen based life due to 1\% and 5\% changes in $m_q^{}$
(in terms of $\bar{A}_s^{}$ and $\bar{A}_t^{}$). To be
precise, for a 5\% change in $m_q^{}$, $\bar{A}_t^{}$ must assume
values within the red (narrow) band to allow for sufficient production of
carbon and oxygen. The most up-to-date
knowledge of these parameters is depicted by the data point with horizontal and vertical error 
bars. This
NNLO determination of $\bar A_{s,t}^{}$ shows that carbon-based life
survives at least a $\simeq 0.7$\% shift in $m_q^{}$. In addition to this
"worst-case scenario", we find that the theoretical uncertainty
in $\bar A_{s,t}^{}$ is also compatible with a vanishing $\partial
\varepsilon / \partial M_\pi^{}$ (complete lack of fine-tuning). Given
the central values of $\bar A_{s,t}^{}$, we conclude that variations of the
light quark masses of $2-3$\% are unlikely to be catastrophic to the
formation of life-essential carbon and oxygen. 


We may also compute the corresponding changes induced by
variations of the EM fine-structure constant $\alpha_{\rm em}^{}$. 
On the lattice, the EM shift receives contributions from the long-range 
Coulomb force and a short-range proton-proton contact interaction. 
The latter contains an unknown coupling strength, which allows for the
regularization of QED on the lattice. 
We have fixed its finite part from the known EM
contribution to the $\alpha$--particle binding energy. 
The dependence of the $E_i^{}$ on $\alpha_{\rm em}^{}$ 
can then be calculated.
By expressing the EM shifts 
as $(\partial X / \partial \alpha_{\rm em}^{})|_{\alpha_{\rm em}^{\rm ph}} \simeq
Q(X)/\alpha_{\rm em}^{}$, we find $Q(\Delta E_b^{}) = 1.19(8)$~MeV,
$Q(\Delta E_h^{}) = 2.80(10)$~MeV and $Q(\varepsilon) = 3.99(9)$~MeV. 
For fixed $m_q^{}$, a variation of $\alpha_{\rm em}^{}$
by $\pm 100~\mathrm{keV} / Q(\varepsilon) \approx 2.5$\% would thus be compatible with 
the formation of carbon and oxygen in our Universe.  
This is consistent with the $\simeq 4$\% bound reported in Ref.~\cite{Oberhummer:2000mn}.

In summary, we have presented \textit{ab initio} lattice calculations of the 
dependence of the triple-alpha process upon the light quark masses and the EM 
fine structure constant. The position of the $^8$Be ground state relative to the two--$\alpha$ 
threshold, as well as that of the Hoyle state relative to the three--$\alpha$ threshold, appears strongly 
correlated with the binding energy of the $\alpha$--particle.  We also find that the formation of 
carbon and oxygen in our Universe would survive a change of $\simeq 2$\% in 
$m_q^{}$
or $\simeq 2$\% in 
$\alpha_{\rm em}^{}$.
Beyond such relatively small changes,
the anthropic principle appears necessary at this time 
to explain the observed reaction rate of the triple-alpha process. In order
to make more definitive statements about carbon and oxygen production
for larger changes in the fundamental parameters, a more precise
determination of $\bar{A}_s^{}$ and $\bar{A}_t^{}$ is needed from future lattice
QCD simulations.


\subsection*{Acknowledgments}
We are grateful to Silas Beane for useful comments and a careful reading 
of the manuscript. We thank Andreas Nogga for providing 
an updated
analysis of the $^4$He nucleus.  Partial financial support from the Deutsche
Forschungsgemeinschaft (Sino-German CRC 110), the Helmholtz Association (Contract No.\ VH-VI-417), 
BMBF (Grant No.\ 06BN9006), and the U.S. Department of
Energy (DE-FG02-03ER41260) is acknowledged.
This work was further supported
by the EU HadronPhysics3 project, and funds provided by the ERC Project No.\ 259218 NUCLEAREFT.
The computational resources 
were provided by the J\"{u}lich
Supercomputing Centre at the Forschungszentrum J\"{u}lich and by
RWTH Aachen. \\

\end{document}